\documentclass[english,12pt,a4paper]{article}
\pdfoutput=1
\usepackage{jheppub}

\usepackage[T1]{fontenc}
\usepackage[utf8]{inputenc}
\usepackage{lmodern, amsmath, amssymb, xcolor, slashed, xfrac}
\usepackage{subfigure}
\usepackage{placeins}
\usepackage{bm}
\usepackage{scalefnt}
\usepackage{comment}
\usepackage{latexsym}
\usepackage{relsize}
\usepackage[normalem]{ulem}
\usepackage{booktabs} 
\usepackage{colortbl} 
\usepackage{mathptmx} 
\DeclareMathAlphabet{\mathcal}{OMS}{cmsy}{m}{n} 
\graphicspath{{figs/}}

\def\figureautorefname~#1\null{Fig.\,#1\null}

\newcommand{\subappref}[1]{\hyperref[#1]{appendix~\ref{#1}}}

\def\equationautorefname~#1\null{Eq.\,(#1)\null}
\renewcommand{\eqref}[1]{\hyperref[#1]{(\ref{#1})}}

\definecolor{light_blue}{rgb}{0.15, 0.35, 0.95}
\definecolor{kit_green}{rgb}{0, 
0.58823 
, 0.50980 
}

\hypersetup{
  colorlinks = true,
  linkcolor = light_blue, 
  citecolor = light_blue, 
  urlcolor = kit_green 
}

\allowdisplaybreaks[4]
\def\beq#1\eeq{\begin{align}#1\end{align}}
\newcommand{\beqa}{\begin{eqnarray}}
\newcommand{\eeqa}{\end{eqnarray}}

\makeatletter
\def\EE{\@ifnextchar-{\@@EE}{\@EE}}
\def\@EE#1{\ifnum#1=1 \times10 \else \times10^{#1}\fi}
\def\@@EE#1#2{\times10^{-#2}}
\makeatother

\newcommand\unit[1]{\,\,\mathrm{#1}}

\newcommand\GeV{\unit{GeV}}
\newcommand\TeV{\unit{TeV}}

\newcommand\iab{\unit{ab^{-1}}}
\newcommand\ifb{\unit{fb^{-1}}}

\def\Babar{{\mbox{\slshape B\kern-0.1em{\smaller A}\kern-0.1em B\kern-0.1em{\smaller A\kern-0.2em R}}}}
\newcommand{\tr}{_{\mathrm{T}}}
\newcommand{\nubar}{\overline{\nu}}

\RequirePackage{xspace}
\def\Bbar    {\kern 0.18em\overline{\kern -0.18em B}{}\xspace}

\newcommand{\ctext}[1]{\raise0.15ex\hbox{\textcircled{\scriptsize{#1}}}}

\makeatletter\g@addto@macro\bfseries{\boldmath}\makeatother
\preprint{P3H--22--018, TTP22--010}
\title{Towards  ruling out the charged Higgs interpretation of the $R_{D^{(*)}}$ anomaly}

\author[a,b]{Monika Blanke,}
\author[a,b]{Syuhei Iguro,}
\author[a]{Hantian Zhang}

\affiliation[a]{Institute for Theoretical Particle Physics (TTP), Karlsruhe Institute of Technology (KIT),
Engesserstra{\ss}e 7, 76131 Karlsruhe, Germany}
\affiliation[b]{Institute for Astroparticle Physics (IAP),
Karlsruhe Institute of Technology (KIT), 
Hermann-von-Helmholtz-Platz 1, 76344 Eggenstein-Leopoldshafen, Germany}

\emailAdd{monika.blanke@kit.edu}
\emailAdd{igurosyuhei@gmail.com}
\emailAdd{hantian.zhang@kit.edu}

\abstract{
Motivated by the notorious anomaly in the lepton flavor universality ratios $R_{D^{(*)}}$, 
we study the sensitivity of  the Large Hadron Collider (LHC) to a low-mass charged Higgs boson $H^-$ lighter than $400\,$GeV in a generic two Higgs doublet model. 
A combination of current constraints from the $B_c\to\tau \nu$ decay, $B_{s}$ meson mixing data, tau sleptons and di-jet searches at the LHC allows to explain the $R_{D^{(*)}}$ anomaly at the $1\,\sigma$ level by a low-mass charged Higgs.
In this context, we estimate the reach of an LHC search for resonant $H^-$ production, where the final state contains an energetic $\tau$ lepton decaying hadronically, a neutrino with large transverse momentum, and an additional $b$-jet ($pp \to b+\tau_h + \nu$). 
Requiring the additional $b$-tagged jet in the $\tau\nu$ resonance search profits from the suppression of the Standard Model background, and therefore it allows us to judge the low-mass $H^-$ interpretation of the $R_{D^{(*)}}$ anomaly.
To demonstrate this, we perform a fast collider simulation for the $\tau \nu$ resonance
search with an additional $b$-tagged jet, and find that most of the interesting parameter region of the whole mass range can already be  probed with the current integrated luminosity of 139$\ifb$.}

\begin{document}

\sloppy 
\maketitle

\renewcommand{\thefootnote}{\#\arabic{footnote}}
\setcounter{footnote}{0}

\section{\boldmath Introduction}
\label{Sec:introduction}
It has been almost a decade since the BaBar collaboration released the astonishing $4\,\sigma$ discrepancy in the lepton universality ratios \cite{Lees:2012xj}
\begin{equation}
    R_{D^{(\ast)}}=\frac{\text{BR}(\overline{B}\to D^{(*)}\tau\nubar)}{\text{BR}(\overline{B}\to D^{(*)}\ell\nubar)}\,,
\end{equation}  
where $\ell = e,\mu$. Since then tremendous  progress has been made to reduce the  theoretical and experimental uncertainties \cite{Lees:2013uzd,Huschle:2015rga,Sato:2016svk,Hirose:2016wfn,Abdesselam:2019dgh,Aaij:2015yra,Aaij:2017deq,MILC:2015uhg,Bernlochner:2017jka,Na:2015kha,Fajfer:2012vx,Bigi:2016mdz,Gambino:2019sif,Aoki:2021kgd,Iguro:2020cpg,Bordone:2019vic}.
On the experimental side, Belle and LHCb have joined the game and the $R_{D^{(\ast)}}$ HFLAV world average \cite{Aoki:2021kgd} has moved towards the SM prediction and the uncertainties have been reduced considerably.
In the meantime, on the theory side, the simple CLN parametrization \cite{Caprini:1997mu} of the $B\to D^{(*)}$ transition form factor has been shown to be insufficient \cite{Bigi:2017njr,Iguro:2020cpg} and the more general parametrization based on heavy-quark effective theory has been proposed up to $\mathcal{O}(\Lambda_{\rm{QCD}}^2/m_c^2)$ \cite{Bordone:2019vic,Iguro:2020cpg}.
As a result, the current significance of the anomaly is about 4$\sigma$ \cite{Iguro:2020cpg}.
Furthermore, more modest but interesting deviations have been observed in the $D^*$ polarization data in $\overline{B}\to D^{*}\tau\nubar$, $F_L^{D^*}$ \cite{Belle:2019ewo}, and in $B_c\to J/\psi \tau\bar\nu$ \cite{LHCb:2017vlu}.
On the other hand, the LHCb collaboration recently reported $R_{\Lambda_c}$=BR$(\Lambda_b\to\Lambda_c\tau\nu)$/BR$(\Lambda_b\to\Lambda_c\mu\nu)$ that is below, albeit consistent with the SM prediction \cite{LHCb:2022piu}. While a suppression of $R_{\Lambda_c}$ below its SM value would rule out a new physics (NP) origin of the $R_{D^{(\ast)}}$ anomaly based on a model-independent sum rule \cite{Blanke:2018yud,Blanke:2019qrx},
 the experimental uncertainty in $R_{\Lambda_c}$ is still too large to draw a clear-cut conclusion.

In this paper we investigate the LHC sensitivity to the low-mass charged Higgs $H^-$ interpretation of the anomaly focusing on a specific channel: final states with an energetic $\tau$ lepton that decays hadronically, 
large missing transverse momentum from an energetic neutrino, and an additional $b$-jet ($pp \to b+\tau_h + E_{\rm T}^{\rm miss} $).
The model has widely been discussed in the literature \cite{Crivellin:2012ye,Crivellin:2013wna,Cline:2015lqp,Crivellin:2015hha,Lee:2017kbi,Iguro:2017ysu,Iguro:2018qzf,Martinez:2018ynq,Fraser:2018aqj,Cardozo:2020uol,Athron:2021jyv}, and recently been revisited in Ref. \cite{Iguro:2022uzz} since the constraint from $B_c\to\tau\nu$ is significantly relaxed \cite{Blanke:2018yud,Blanke:2019qrx,Aebischer:2021ilm}.
The revision \cite{Iguro:2022uzz} found that a charged scalar can still explain the $R_{D^{(\ast)}}$ anomaly within the 1\,$\sigma$ region when $m_{H^-}\le 400\,$GeV holds.
It is noted that the charged-Higgs scenario with larger mass is excluded by the $\tau\nu$ resonance search at the LHC \cite{Iguro:2018fni,Sirunyan:2018lbg} \footnote{Since the search for low-mass $H^-$ suffers from the huge SM background from the $W$-boson tail, the LHC Run 2 data have not been interpreted for  $m_{H^-}\le400$ GeV.}, and the low-mass bottom flavored di-jet search \cite{CMS:2018kcg,ATLAS:2019itm} and a conventional search for tau sleptons \cite{CMS:2021woq} constrain the available parameter region.
The result clearly shows the importance of the improvement in $\tau\nu$ resonance searches \cite{Iguro:2022uzz}, which is the main subject of this paper.
From the results obtained in Refs.\,\cite{Altmannshofer:2017poe,Iguro:2017ysu,Abdullah:2018ets,Marzocca:2020ueu,Iguro:2020keo,Endo:2021lhi}, one can infer that requiring an additional $b$-tagged jet is also effective in probing the low mass window. The reason is that the additional $b$-jet reduces the number of SM-originated background (SM BG) events and thereby improves the signal to BG ratio.
However this technique has not yet been used in the experimental analyses.
In this paper, we will thus employ this technique and quantify its impact on the LHC sensitivity to a low-mass charged Higgs boson.

The rest of the paper is structured as follows.
A simplified $H^-$ model and its relevant parameters are introduced in Sec.~\ref{Sec:model}.
We propose an LHC search strategy for the $b\tau\nu$ signature in Sec.~\ref{Sec:LHC} including the relevant kinematic cuts, and describe our method in generating signal and background events.
The resulting collider prospects and their impact on the $H^-$ interpretation of the $R_{D^{(*)}}$ anomaly are discussed in Sec.~\ref{Sec:result}.
Finally Sec.~\ref{Sec:conclusion} is devoted to the conclusions.

\section{Model and parameters}
\label{Sec:model}

We now introduce the simplified model for a charged scalar boson $H^-$ solving the $R_{D^{(\ast)}}$ anomaly.
Such a charged Higgs emerges from the second SU(2) doublet of a generic two Higgs doublet model (G2HDM), along with CP even and odd neutral scalars.
In the model under consideration the additional Higgs doublet couples to all fermions, a setup which appears in many UV models, such as the left-right model \cite{Ball:1999mb,Kiers:2002cz,Zhang:2007da,Guadagnoli:2010sd,Blanke:2011ry,Frank:2011jia,Bertolini:2014sua,Bernard:2015boz,FileviezPerez:2017zwm,Iguro:2018oou,Iguro:2021nhf} and even in the TeV scale Pati-Salam model to break the symmetry with a bi-doublet field \cite{Iguro:2021kdw,Iguro:2022ozl}. 
In general such a coupling structure is dangerous since the additional neutral scalars possess flavor violating interactions at tree level. \cite{Atwood:1996vj}.
A detailed analysis of the model's flavor phenomenology can be found in Refs. \cite{Crivellin:2013wna,Iguro:2017ysu}.

Following Ref. \cite{Iguro:2022uzz}, we introduce the simplified interaction Lagrangian for a charged scalar $H^-$ entering $R_{D^{(*)}}$ as
\begin{align}
{\cal L}_{int}=
& + y_Q H^- (\overline{b} P_R c)
- y_\tau H^- (\overline{\tau} P_L \nu_{\tau})   +{\rm{h.c.}},\label{Eq:G2HDM}
\end{align}
and we focus on the low-mass window 
\begin{align}
180\,\text{GeV}\le m_{H^-}\le 400\,\text{GeV}, 
\end{align}
which is currently not constrained by direct searches at colliders.
By integrating out the heavy degrees of freedom, the low-energy effective Hamiltonian describing $b\to c\tau\nu$ transitions is given as
\begin{align}
 {\mathcal H}_{\rm{eff}}= 
 2 \sqrt 2 G_F V_{cb} \Bigl[ 
 & (\overline{c} \gamma^\mu P_Lb)(\overline{\tau} \gamma_\mu P_L \nu_{\tau}) +C_{S_L}(\overline{c} P_L b)(\overline{\tau} P_L \nu_{\tau}) \Bigl{]},\label{Eq:Hamiltonian}
\end{align} 
with $P_{L/R}=(1\mp\gamma_5)/2$ being the chirality projection operators.
The first term corresponds to the SM contribution, stemming from a tree-level $W^-$ exchange.
With the  normalization fixed in Eq.\,(\ref{Eq:Hamiltonian}), we have $C_{S_L}=y_{Q}^* y_\tau/m_{H^-}^2/(2 \sqrt 2 G_F V_{cb})$ and use $V_{cb}=0.042$ hereafter.
In addition, we employ the numerical description of $R_D$, $R_{D^*}$ and BR($B_c\to\tau\nu$) given in \cite{Blanke:2018yud},\footnote{Similar numerical formulae can be found in Ref.\,\cite{Iguro:2018vqb}.}
\begin{align}
R_D&\simeq R_D^{SM}\biggl{(}1+1.54{\rm Re}\bigl[ C_{S_L}\bigl]+1.09| C_{S_L}|^2\biggl{)},\\
R_{D^*}&\simeq R_{D^*}^{SM}\biggl{(}1-0.13{\rm Re}\bigl[ C_{S_L}]+0.05|C_{S_L}|^2\biggl{)},\\
\rm{BR}&(B_c\to\tau\nu)\simeq 0.02|1-4.3C_{S_L}|^2,
\label{Eq:RDS}
\end{align}
where the Wilson coefficient (WC) \,$C_{S_L}$ is defined at the $b$-quark mass scale of $m_b=4.2$\,GeV. 

In this work we restrict ourselves to the scenario in which only the couplings $y_Q$ and $y_\tau$ are nonzero. The Yukawa term  $y_{Q_d}H^+ (\bar c P_Rb)$ is severely constrained by $B_s$ mixing mediated by the neutral scalars, thus it is difficult to significantly enhance $R_{D^{(*)}}$ with this coupling.
Other Yukawa-originated contributions to $b\to c\tau\nu$  receive a CKM suppression in the WC defined in Eq.\,(\ref{Eq:Hamiltonian}), and hence their impact on $R_{D^{(*)}}$ is small.
Phenomenologically our choice is a good approximation since other couplings are constrained when $y_Q$ and $y_\tau$ are sizable.
For instance additional Yukawa couplings to  light leptons are stringently constrained by $B_c\to e\nu$ and $B_c\to \mu\nu$, due to the even larger chirality enhancement factor.
An additional top Yukawa coupling is constrained by ${B_s}$ mixing data and the heavy neutral Higgs search in a tauonic decay \cite{ATLAS:2020zms}, see fig. 13 of Ref.\,\cite{Iguro:2017ysu}. Furthermore the decay $H^-\to \bar{t} b$ induced by the top Yukawa coupling is kinematically suppressed when $m_{H^-}\simeq 200$\,GeV.
Other quark Yukawa couplings especially to the light quark generations are dangerous since they contribute to heavy Higgs production and therefore have to be suppressed.
Therefore we conclude that allowing only the couplings $y_Q$ and $y_\tau$ to be nonzero is sufficient for the purpose of our analysis, since the presence of other couplings cannot significantly affect the $b\tau\nu$ event number and it does not worsen the sensitivity. 

\section{Collider phenomenology}
\label{Sec:LHC}

According to the findings in Ref. \cite{Iguro:2022uzz}, the charged-Higgs solution to the  $R_{D^{(*)}}$ anomaly can directly be probed at the LHC, once the experimental sensitivity to low-mass charged Higgs bosons in $\tau\nu$ resonance searches is improved. In this section we discuss the strategy to achieve this goal, as well as our procedure for the generation of signal and relevant SM BG events.

\subsection{Strategy}
\label{Sec:strategy}

Currently experimental constraints on resonant $H^-$ production decaying to $\tau\nu$ are available from LHC Run 2 for $m_{H^-}\ge400\,$GeV \cite{Sirunyan:2018lbg} and $m_{H^-}\ge500\,$GeV \cite{ATLAS:2021bjk}, and from LHC Run 1 for $m_{H^-}\ge300\,$GeV \cite{CMS:2015hmx}.
These analyses originally searched for a $W^\prime$ boson in a sequential standard model, looking for a single hadronically decaying $\tau$ lepton in association with missing transverse energy.
The key kinematic variable discriminating the signal from the SM BG is the transverse mass,  
\begin{align}
 m\tr = \sqrt{2p\tr^{\tau_h} E\tr^\textrm{miss} \left[1- \cos \Delta \phi (\vec{p}\tr^{\tau_h}, \vec{p}\tr^{\,\rm miss})\right]}  \,,
\end{align} 
where $\Delta \phi$ is the relative angle between the two momenta $(0\leq \Delta \phi\leq \pi)$, and the missing transverse momentum is expressed by $ \vec{p}\tr^{\,\rm miss}$ with magnitude $E_{\rm T}^{\rm miss}$.
$\tau_h$ stands for the hadronic objects from the $\tau$ decay.
The low $m_{\rm{T}}$ region suffers from the huge SM BG which stems from the tail of the $W$ boson. The latter is dominantly produced through
\begin{align}
u\bar{d}/d\bar{u}\to W^+ /W^- \to \tau^+ \nu/\tau^- \nu. \nonumber
\end{align}
It is worth noting that since $u$ and $d$ quarks can be valence quarks, there is a charge asymmetry in the number of $W^+ /W^-$ bosons produced.
However the $W$ resonance is not heavy, so that the sea quark contribution can be sizable, diluting the asymmetry.
For a charged Higgs resonance with the coupling structure defined in Eq.\,(\ref{Eq:G2HDM}), the initial state does not involve the $u,d$ valence quarks, and hence $H^+$ and $H^-$ are produced at equal rates. On the other hand in the present paper we are interested in the low-mass region, $m_{H^-}\le 400\,\text{GeV}$, for which the sea quark contribution is more relevant. Furthermore, since we require an additional $b$-tagged jet, as discussed below, the main SM background processes are single-top and $t\bar t$, so that the charge asymmetry of the SM background is expected to be much less pronounced. Consequently, for the sake of simplicity, we will not impose a selection cut based on the charge of the $\tau$ lepton.

We next argue how the requirement of an additional $b$-tagged jet in the final state can further improve the sensitivity of the charged Higgs searches. The importance of such a $b$-jet requirement was first realized in  Ref.\,\cite{Altmannshofer:2017poe}, using a reference NP scale of  $m_{\text{NP}}=1$~TeV.
Ref.\,\cite{Iguro:2017ysu} demonstrated the impact of the additional $b$-jet in a parton-level comparison performed within the G2HDM but fixing $m_{H^-}=500$ GeV for simplicity.
Including a fast detector simulation, Ref.\,\cite{Abdullah:2018ets} showed that an additional flavor tagging is useful to search for low-mass $W^\prime$ scenarios. 

\begin{figure}[t]
\begin{center}
\includegraphics[scale=0.5]{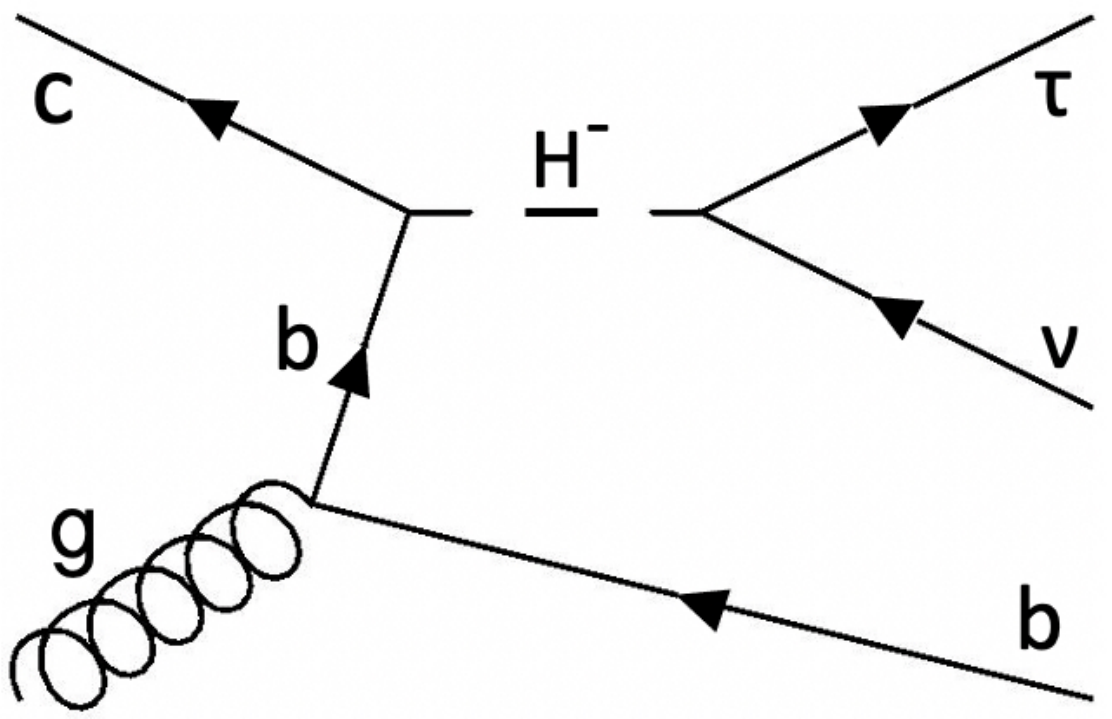}~~~~~~~~~~~~~
\includegraphics[scale=0.415]{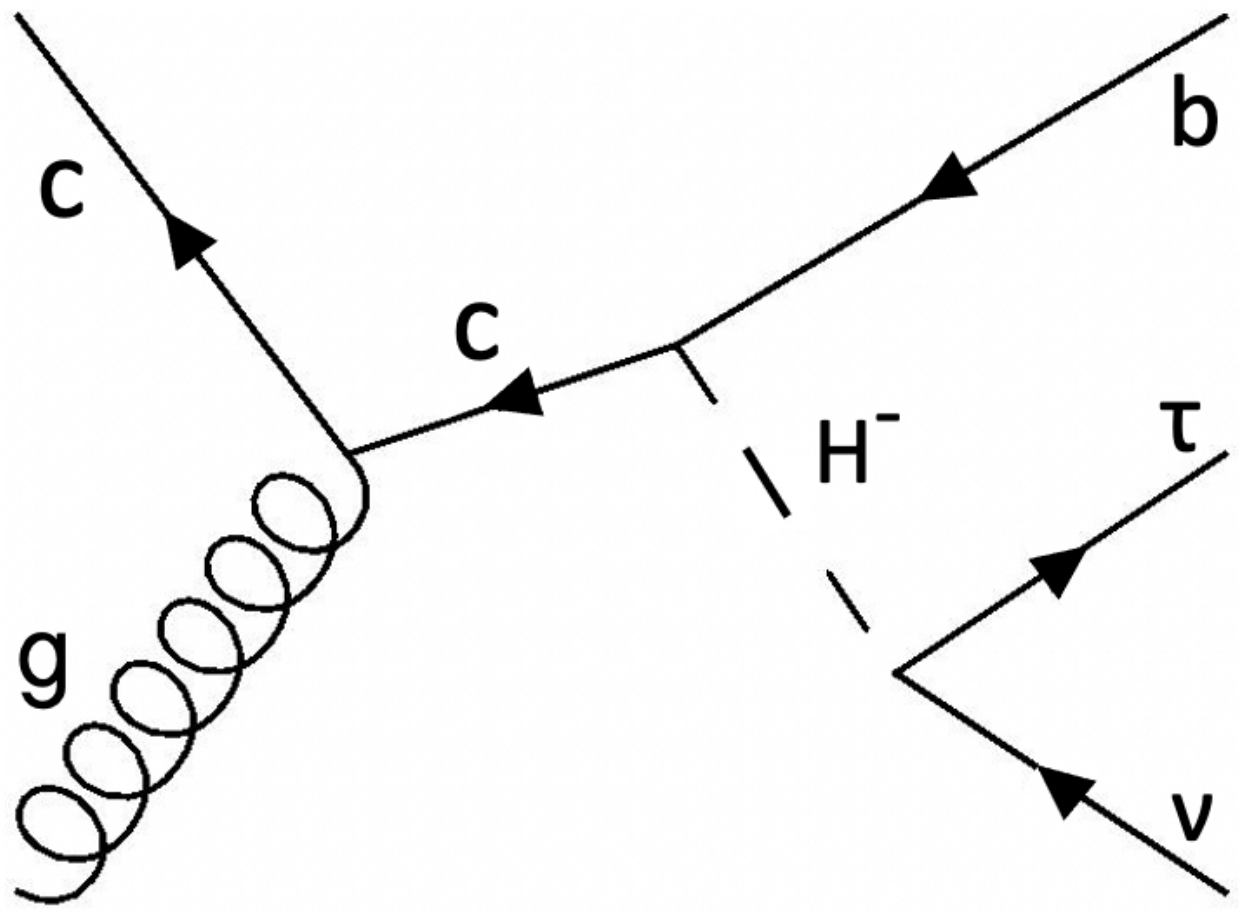}
\caption{
\label{Fig:Dia}
Representative Feynman diagrams for the $b\tau\nu$ signal from $H^-$.
} 
\end{center}
\end{figure}

An additional $b$-tagging is effective to further reduce the SM BG to our $H^-$ resonance search, since the  process
\begin{align}
u_i g  \to b W \to b \tau \nu
\end{align}
receives a suppression factor $|V_{u_ib}|^2$ ($u_i = u,c$),
while the pollution from 
\begin{align}
qg \to  j W \to j \tau \nu 
\end{align}
is suppressed by the mis-tagging rate $\epsilon_{j\to b}$ to meet the $b$-tagging requirement. The $b$-tag requirement also serves to efficiently suppress the ``fake $\tau$'' BG from QCD jets.
As a result single top and $t \bar{t}$ constitute the dominant BGs in the signal region.
Representative Feynman diagrams contributing to the $b\tau\nu$ signature from $b$-jet-associated charged-Higgs production  are shown in Fig.\,\ref{Fig:Dia}.

In our model the heavy resonance $H^-$ only couples to the third generation fermions and the charm quark, and also the spin structures of the $H^-$ and the $W^\prime$ are different. Hence, the efficiency and the acceptance of the selection cuts need to be estimated by a Monte Carlo simulation.

The signal cross-section in our model can be parameterized in terms of $(m_{H^-},\,y_Q,\, y_\tau)$ as follows:
\begin{align}
\sigma (pp \to H^- +b) \times {\rm{BR}}(H^- \to \tau \nu) 
= \sigma_0(m_{H^-}) \times \frac{|y_Q|^2 |y_\tau|^2}{3|y_Q|^2 +|y_\tau|^2},
\label{Eq:Sig_Xs}
\end{align}
where $\sigma_0(m_{H^-})$ is a function of the charged Higgs mass only, while the $y_Q$ dependence has been factorized. 
Note that flavor physics constraints, e.g. from $B$ meson mixings, preclude large $y_Q$ values, and thus the narrow width approximation is viable.

\subsection{\boldmath Event generation}
\label{Sec:event_generation}

Both NP signal and SM BG processes are simulated with Monte Carlo (MC) event generators at $\sqrt{s}=13\TeV$.
Event samples generated using {\sc\small MadGraph}5\_a{\sc\small MC}@{\sc\small NLO}~v3.2.0 \cite{Alwall:2014hca} 
are interfaced with {\sc\small PYTHIA} v8.3~\cite{Sjostrand:2014zea} for hadronization and decay of the partons.
{\sc\small NNPDF}2.3 \cite{Ball:2012cx} in the five-flavor scheme is adopted and the MLM merging is used \cite{Alwall:2007fs}.
Detector effects are simulated based on {\sc\small Delphes} v3.4~\cite{deFavereau:2013fsa}. 
Jets are reconstructed using the anti-$k_T$ algorithm~\cite{Cacciari:2008gp} with a radius parameter of $R=0.5$.

\subsubsection{\boldmath Background simulation}
\begin{table}[t]
\centering
  \scalebox{1.08}{
\begin{tabular}{c}\hline
selection criteria\\ \hline\hline
$N_b=1$, $p_T^b \geq 30\GeV$, $|\eta_b| < 2.5$\\ 
$N_{\tau_h}=1$, $p_T^{\tau_h}\ge 70\GeV$, $|\eta_{\tau_h}| \le 2.1$\\ 
$E_T^{\text{miss}} \ge 80\GeV$\\
$N_{e,\mu}=0$, $p\tr^{e,\mu}\ge 20\GeV$, $|\eta_e| \le 2.5$ or $|\eta_\mu| \le 2.4$\\
$N_j \leq 2\,$, $p_T^j \geq 20\GeV$, $|\eta_j| \leq 2.5$\\
$\Delta\phi(\vec{p}_T^{\tau_h}, \vec{p}_T^{\,\rm miss}) \ge 2.4$, $0.7 \le p_T^{\tau_h}/E_T^{\rm miss} \le 1.3$\\
\hline\hline\end{tabular}}
\caption{
\label{tab:kincut}
Summary of kinematic cuts.  
}
\end{table}
The SM BG events are generated
following the method explored in Ref.~\cite{Endo:2021lhi}.
Motivated by the previous phenomenological studies and experimental analyses, we consider five BG categories: $Wjj$, $Zjj$, ${t \bar{t}}$, single top, and $\boldsymbol{VV}(=WW,\,ZZ,\,WZ)$.
More explicitly different from Ref.~\cite{Endo:2021lhi}, we combined $Zjj$ with $Z$ or $\gamma$\,(Drell-Yan) categories and renamed them as $Zjj$ for simplicity.
We generated 5M, 15M, 8M, 10M, and 3M events, respectively, for the five BG categories.
A detailed process description is available in section 3.1 of Ref.\,\cite{Endo:2021lhi}.

To study the sensitivity of the $b\tau\nu$ search the following set of kinematic cuts is considered.
We require exactly one $b$-tagged jet with $p_T^b \geq 30\GeV$ and $|\eta_b| < 2.5$, and exactly one $\tau$-tagged jet with the transverse momentum of $\tau_h$ satisfying $p_T^{\tau_h}\ge 70\GeV$, and the pseudo-rapidity of $\tau_h$, $|\eta_{\tau_h}| \le 2.1$. 
We also impose the large missing transverse momentum condition, $E_T^{\text{miss}} \ge 80\GeV$,  to suppress the large $W$ resonance contribution, and we reject events with isolated light leptons with $p\tr^{e,\mu}\ge 20\GeV$ within $|\eta_e| \le 2.5$ or $|\eta_\mu| \le 2.4$.
Furthermore, we restrict the number of light-flavored jets, $N_j \leq 2\,$, to suppress the top-originated backgrounds, where the jets satisfy $p_T^j \geq 20\GeV$ and $|\eta_j| \leq 2.5$.
Then, to select the back-to-back configuration in which the missing momentum is balanced with the $\tau$-tagged jet, we require $\Delta\phi(\vec{p}_T^{\tau_h}, \vec{p}_T^{\,\rm miss}) \ge 2.4$ and $0.7 \le p_T^{\tau_h}/E_T^{\rm miss} \le 1.3$.
Note that in order to focus on the low-mass resonance the $p_T^{\tau_h}$ and $E_T^{\rm{miss}}$ thresholds are lowered compared to the selection cuts in Ref.\,\cite{Endo:2021lhi}.
The above cuts are summarized in Tab.\,\ref{tab:kincut}.

An energetic $\tau$ lepton can also stem from the decay of an energetic hadron, however, it is likely to be accompanied by nearby jets and hence vetoed by $\tau$ isolation criteria.
Therefore we do not consider BG events with $\tau$ whose parent particle is a meson or baryon.
For the $\tau$-tagging efficiency, the ``VLoose'' working point is adopted for the hadronic decays: $\epsilon_{\tau \to \tau}=0.7$~\cite{CMS:2018jrd}.
For the mis-tagging rates, 
we apply $p_T^j$-dependent efficiency based on Ref.~\cite{CMS:2018jrd}.
As a reference the mis-tagging rate $\epsilon_{c,b\to\tau}$ is assumed to be 7.2$\times10^{-4}$.
As for the $b$-tagging efficiencies, the following working point is applied based on Table\,4 of Ref.~\cite{ATLAS:2019bwq}, 
\begin{align}
 &\epsilon_{b\to b}=0.6\,,&
 &\epsilon_{c \to b}=1/27\,,&
 &\epsilon_{j\to b}=1/1300\,.&
\label{Eq:b-tag}
\end{align} 
 
\begin{figure}[t]
\begin{center}
\includegraphics[scale=0.5]{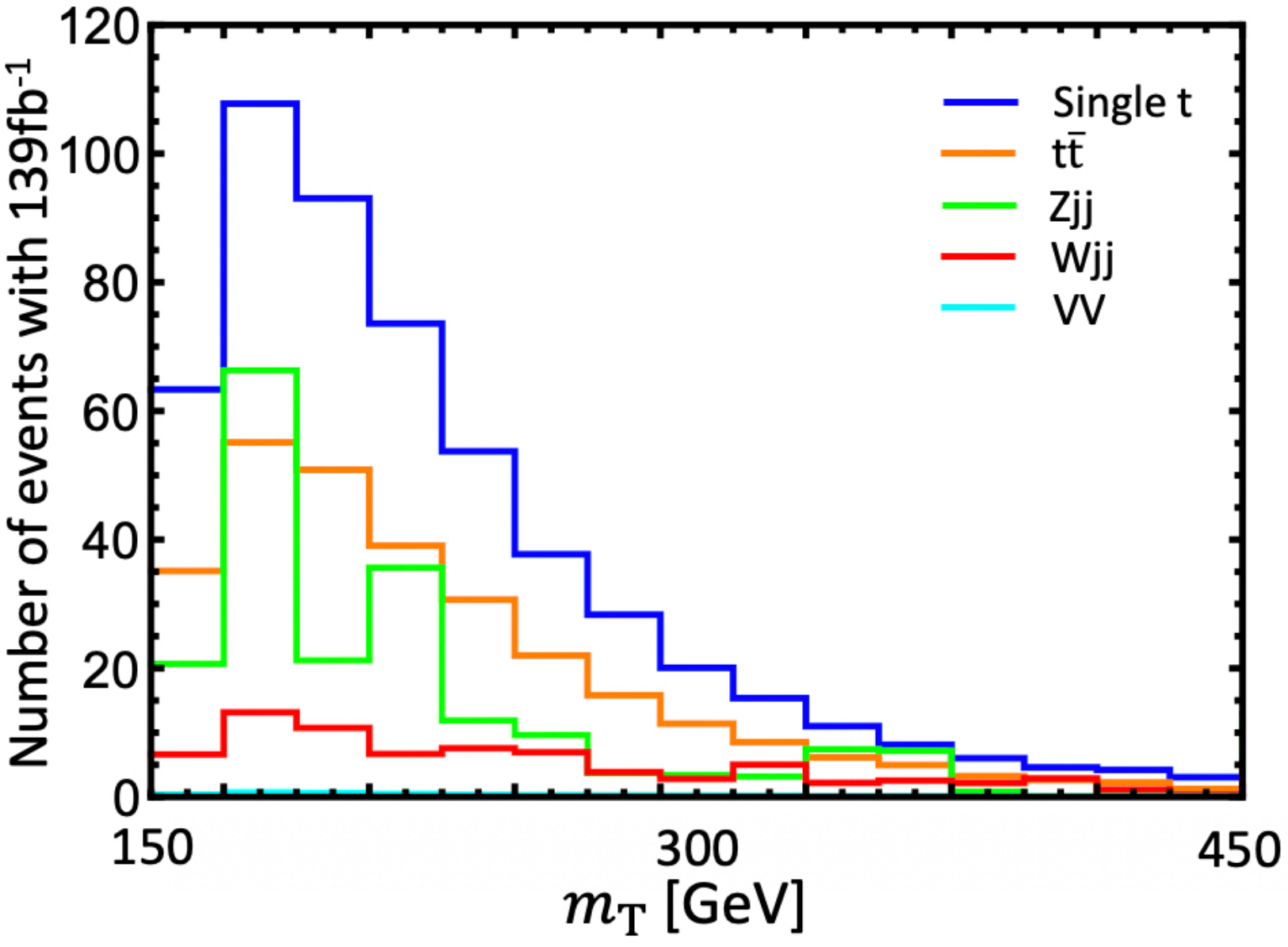}
\caption{
\label{Fig:MT_BG}
Expected SM BG $m_{\rm{T}}$ distributions after applying all kinematic cuts, shown as number of events per $m_{\rm{T}}$ bin.
We assumed 139\,$\ifb$ of data.
The colors distinguish the different BG categories, as detailed in the upper right corner.
} 
\end{center}
\end{figure}

The resulting $m_{\rm{T}}$ distribution of the SM BG after applying the kinematic cuts described above is shown in Fig.\,\ref{Fig:MT_BG} in the range 150\,GeV$\le m_{\rm{T}}\le$450\,GeV. Here we have chosen an $m_{\rm{T}}$ binning with 20\,GeV steps.
This $m_{\rm{T}}$ bin width is moderate as seen in Ref.\,\cite{ATLAS:2021bjk}.
As seen from Fig.\,\ref{Fig:MT_BG}, single top gives the largest BG contribution for the whole $m_{\rm{T}}$ region.
This is mainly due to the cut on the number of light-flavored jets $N_j\le2$, which is not introduced in Ref.\,\cite{Abdullah:2018ets}.
The next-to-leading contribution comes from $t\bar{t}$.
The event distribution of the $Zjj$ category appears statistically unstable even with 15M of simulated events. However, our statistical method which we explain in the next section suppresses the possible bias.

\subsubsection{\boldmath Signal simulation}
\label{Sec:signal}

Within the simplified $H^-$ model of section \ref{Sec:model}, we generate
100K signal events 
for the following set of $H^-$ masses,
\begin{align}
m_{H^-}= \{ 180\,, 200\,,250\,,300\,,350\,,400\,\}\GeV,
\label{Eq:mass_range_collider}
\end{align}
and allowing for up to two additional jets.
In the event generation we set $y_Q=y_\tau=1$ and rescale the signal cross section based on Eq.\,(\ref{Eq:Sig_Xs}).
The width-to-mass ratio of this working point is about $8\%$.
This choice leads to a small dilution of the $m_{\rm{T}}$ distribution, which could result in too conservative sensitivity estimates.
The NP-SM interference is expected to be negligible due to the resonance nature of the signal and the smallness of the SM $pp\to b\tau\nu$ amplitude.

\begin{figure}[t]
\begin{center}
\includegraphics[scale=0.5]{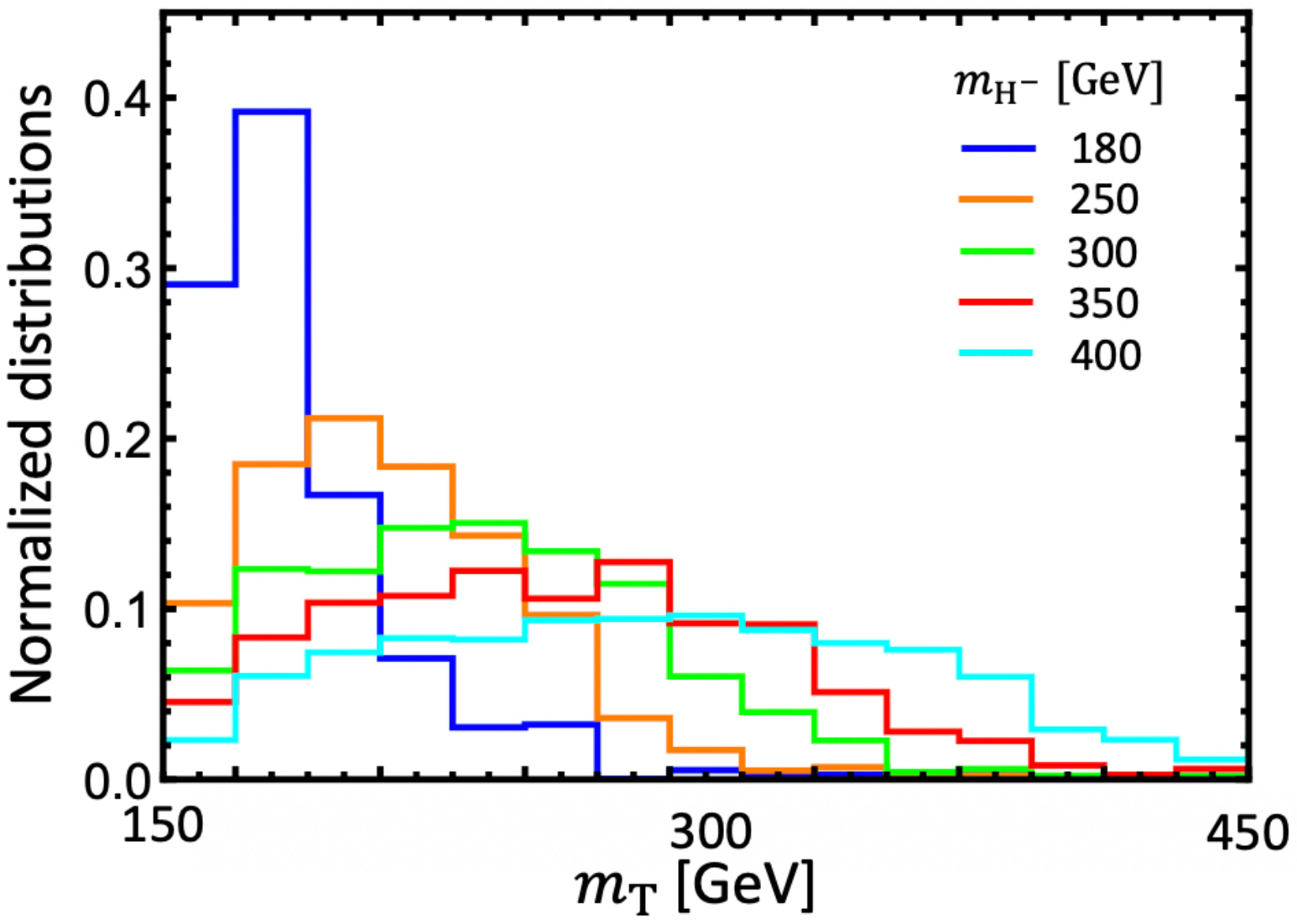}
\caption{
\label{Fig:MT_sig}
Normalized  signal $m_{\rm{T}}$ distributions after imposing all kinematic cuts. The different colors correspond to different charged-Higgs masses, as listed in the upper right corner.
} 
\end{center}
\end{figure}

The representative normalized signal $m_{\rm{T}}$ distribution after imposing the kinematic cuts is shown in Fig.\,\ref{Fig:MT_sig} for the various masses, as detailed in the plot.
It is noted that the $m_{\rm{T}}$ distribution for $m_{H^-}=200\,$GeV is similar to that of $m_{H^-}=180\,$GeV and thus not shown.
The expected signal event numbers after imposing the above kinematic cuts,  in the range $150\,{\rm{GeV}}\le m_{\rm{T}}\le450\,{\rm{GeV}}$, assuming 139$\ifb$, and fixing  $y_Q=y_\tau=1$ with $m_{H^-}=$ 180, 200, 250, 300, 350, and 400\,GeV are $8.2\times10^4$,\,$1.1\times10^5$,\,$1.3\times10^5$,\,$8.5\times10^4$,\,$6.6\times10^4$, and \,$4.7\times10^4$, respectively. Interestingly, the number of signal events after imposing the kinematic cuts varies only mildly with increasing mass $m_{H^-}$, despite the steeply decreasing function $\sigma_0(m_{H^-})$.
This is a direct consequence of the broader $m_T$ distribution.
As a result, the sensitivity of the $b\tau\nu$ search in the $(y_Q,y_\tau)$ coupling plane depends only mildly on $m_{H^-}$, as we will see below.

\section{\boldmath Results}
\label{Sec:result}

We now turn to the discussion of the results for the $b\tau\nu$ search proposed in this paper.
We first quantify the sensitivity of the $b\tau\nu$ signal to a low-mass charged-Higgs boson, using the currently available $139\,\ifb$ of LHC data.
We then discuss the implications for the charged-Higgs solution of the $R_{D^{(*)}}$ anomaly.

\subsection{Sensitivity of $b\tau\nu$ search}
In order to determine the sensitivity of the $b\tau\nu$ signature to a low-mass charged Higgs, we follow the procedure in Ref.\,\cite{Endo:2021lhi}.
To account for statistical uncertainties, we employ Poissonian statistics.

Based on the CMS analysis with 36\,fb$^{-1}$ of data \cite{Sirunyan:2018jdk}, $30\%$ of systematic uncertainty is assigned to the BG as a conservative estimate. In addition, to be conservative, we also assign a $30\%$ systematic uncertainty to the signal, in order to account for PDF and scale uncertainties.

\begin{figure}[p]
\begin{center}
\includegraphics[scale=0.405]{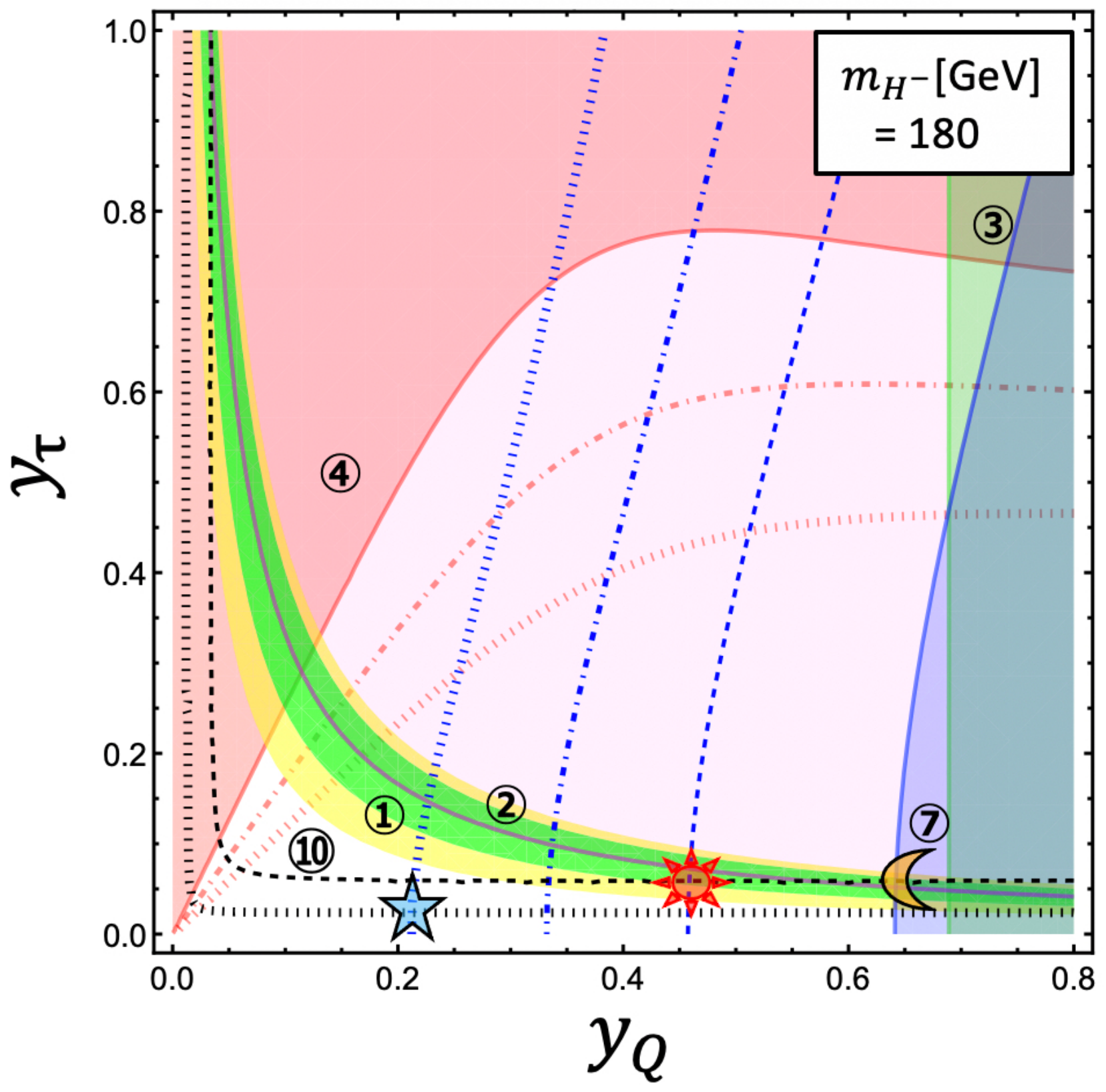}
\includegraphics[scale=0.405]{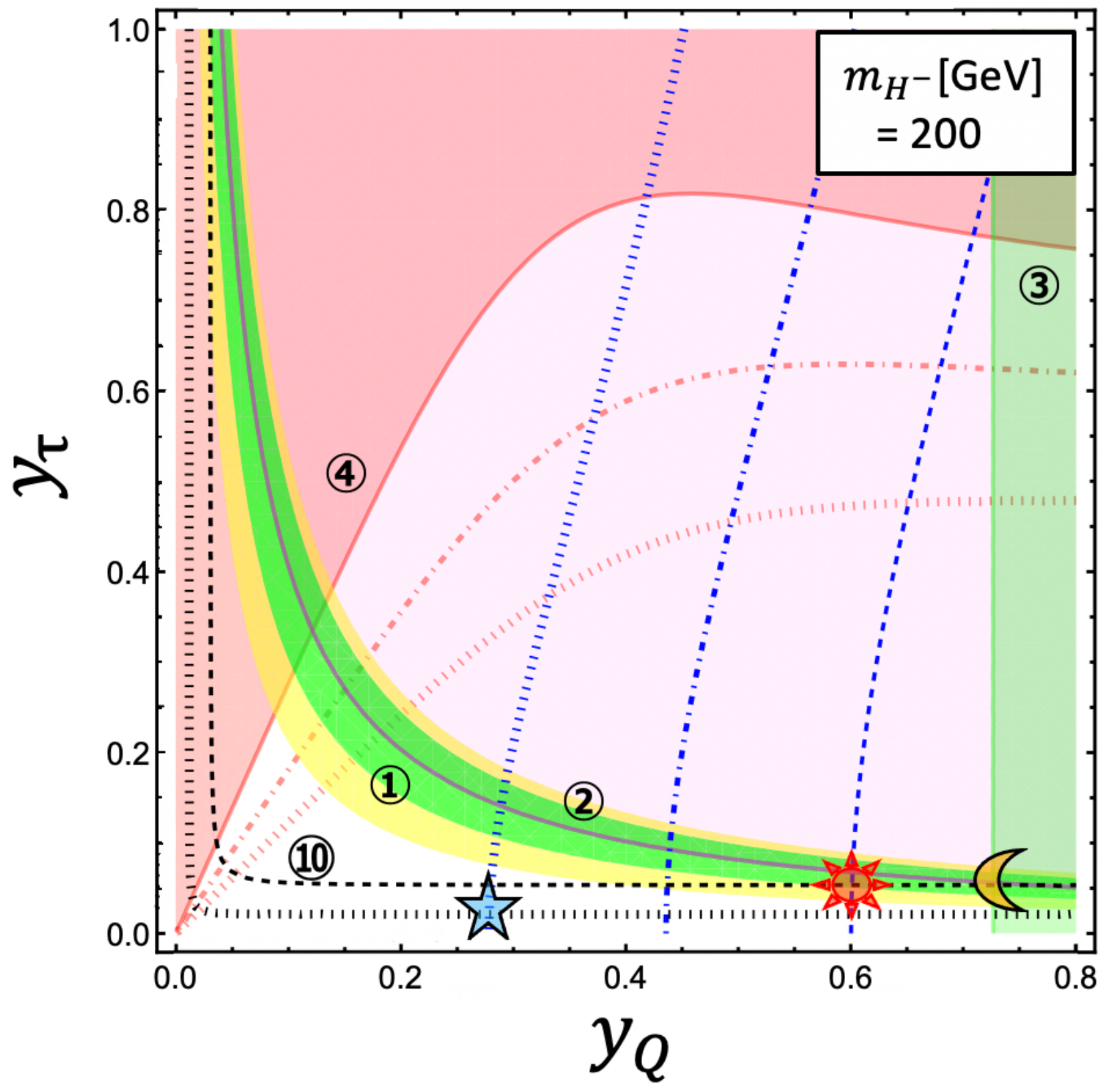}
\includegraphics[scale=0.405]{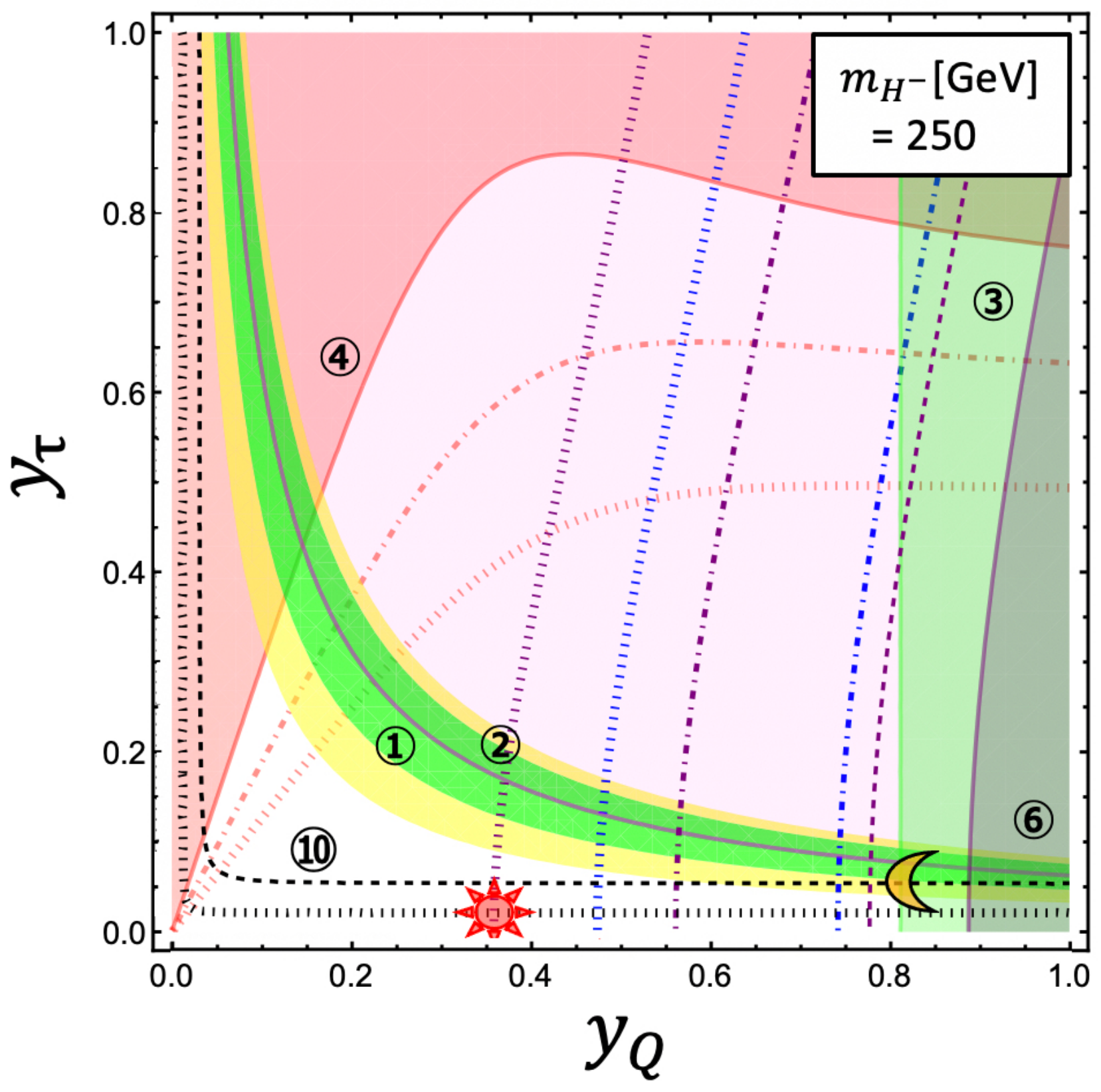}
\includegraphics[scale=0.405]{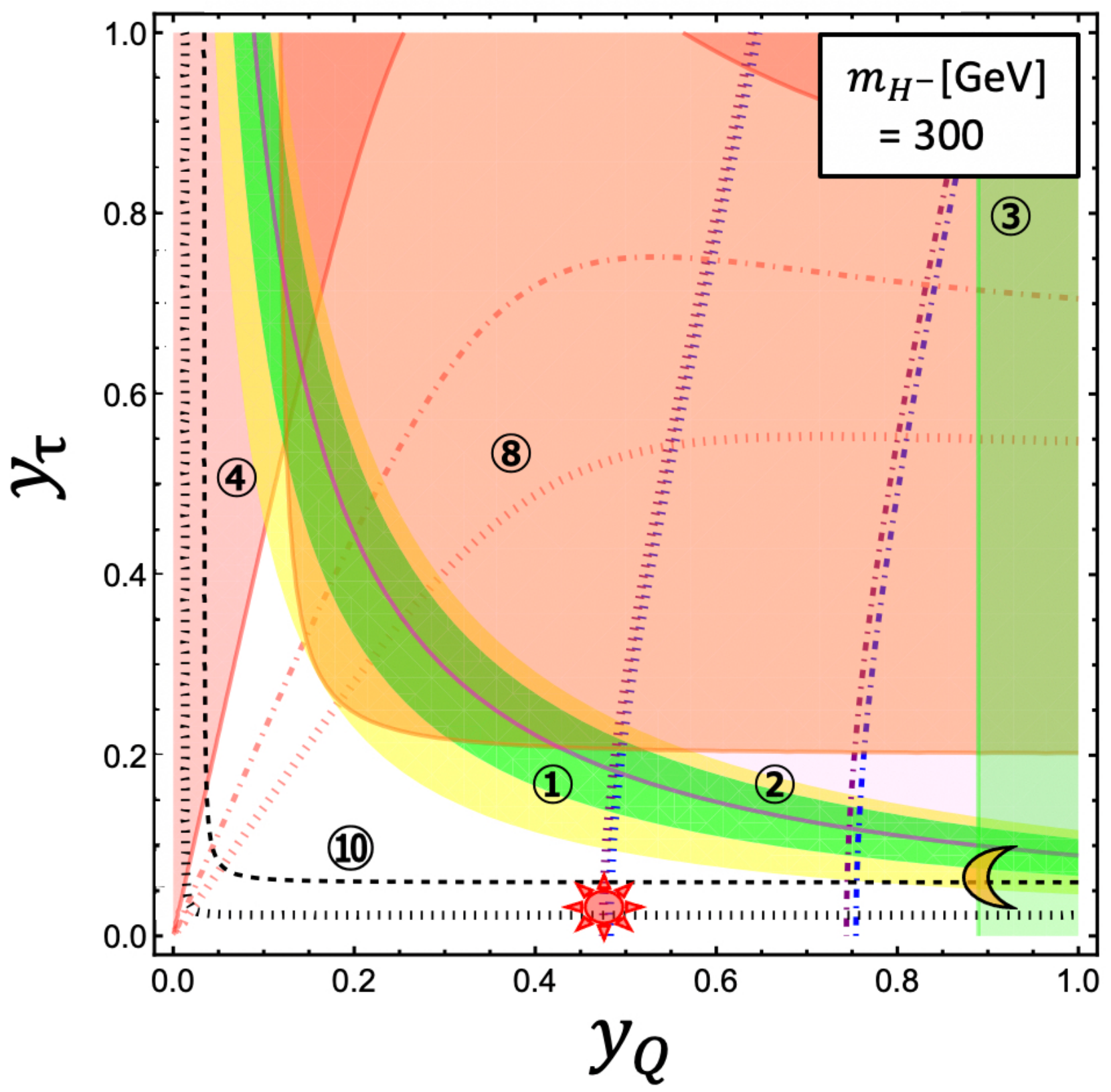}
\includegraphics[scale=0.405]{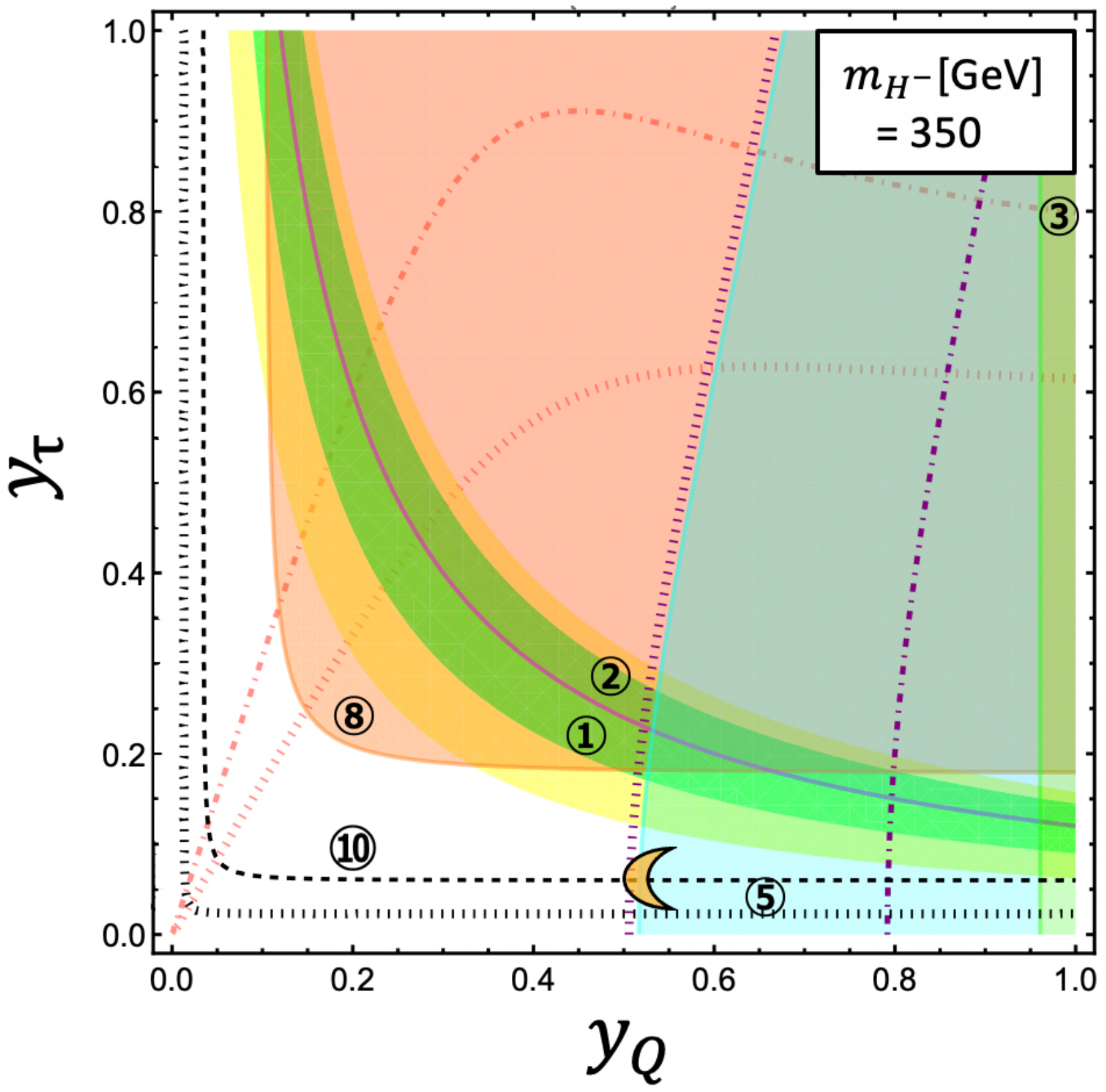}
\includegraphics[scale=0.405]{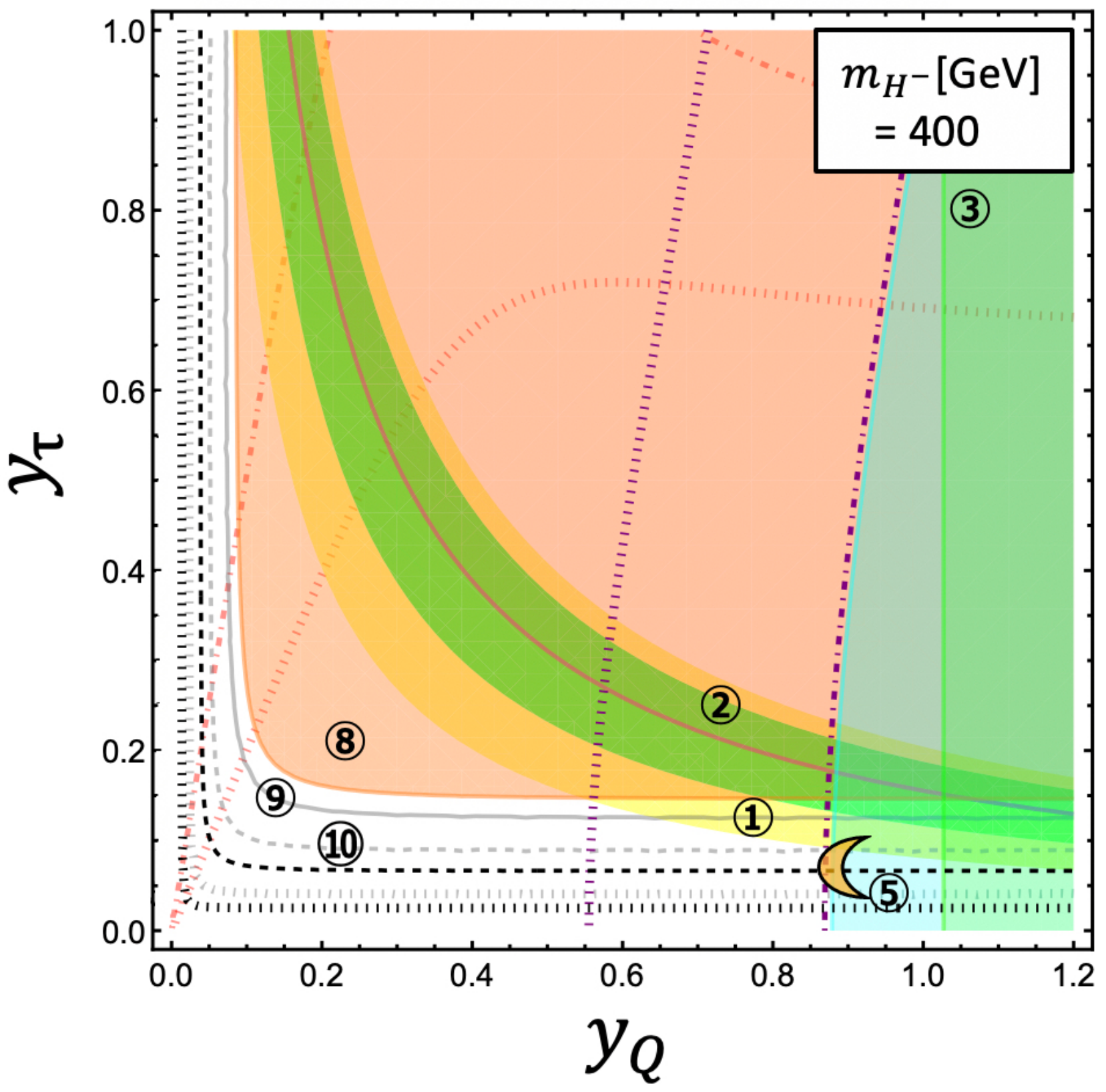}
\caption{
\label{Fig:coupling} 
The sensitivity of the $b\tau\nu$ search channel is shown by dashed black lines in the $(y_Q,y_\tau)$ coupling plane.
The charged Higgs mass is fixed to be $180,\,200,\,250,\,300,\,350,$ and $400$\,GeV, as shown in the upper right corner of the respective panel.
In each panel we define  benchmark points denoted by a crescent moon, sun and star.
The numerical values of the corresponding couplings can be found in the main text.
Other flavor and collider constraints are also shown. 
The circled numbers express the relevant observables and processes defined in Tab.\,\ref{Tab:cosnts}.
Solid lines show the current constraint while dashed, dotted-dashed, dotted lines correspond to the projected sensitivity with $139\,\ifb,\,500\,\ifb,\,$ and $3\,\iab$.
} 
\end{center}
\end{figure}

\begin{table}[t]
\centering
  \scalebox{0.92}{
  \begin{tabular}{c|| c||  c || c || c } 
  \rowcolor{white} 
  \hline\hline
  Process & Couplings & Mass range  & Number, color & Ref.\\  \hline\hline
  $R_{D^{(*)}}$ & $y_Q \times y_\tau$ & all &\ctext{1}, {{\color[rgb]{0,0.7,0} green}}($1\sigma$) and {{\color[rgb]{0.85,0.85,0} yellow}}($2\sigma$)& \cite{Aoki:2021kgd}\\  \hline 
  $B_c\to\tau\nu$ & $y_Q \times y_\tau$ & all &\ctext{2}, {{\color[rgb]{1,0.62,0.68} light pink}}& \cite{Aebischer:2021ilm}\\  \hline
  $B$ meson mixings & $y_Q$& all &\ctext{3}, {{\color[rgb]{0,1,0} light green}}&\cite{DiLuzio:2019jyq}\\  \hline 
  stau search & $y_\tau$ ($y_Q$) & all &\ctext{4}, {{\color[rgb]{0.85,0,0} red}}&\cite{CMS:2021woq}\\  \hline 
  $2b$ & $y_Q$ ($y_\tau$) &$m_{H^-}\ge325\,$GeV&\ctext{5}, {{\color[rgb]{0,1,1} cyan}}&\cite{CMS:2018kcg}\\  \hline
  $2j$ &$y_Q$ ($y_\tau$)&$m_{H^-}\le300\,$GeV&\ctext{6}, {{\color[rgb]{0,0,1} blue}}&\cite{CMS:2017dcz}\\  \hline 
 $2b+\gamma$ &$y_Q$ ($y_\tau$)&$m_{H^-}\ge225\,$GeV&\ctext{7}, {{\color[rgb]{0.4,0,0.4} purple}}&\cite{ATLAS:2019itm}\\  \hline 
  $\tau\nu$ (Run\,1)&$y_Q \times y_\tau$& $m_{H^-}\ge300\,$GeV&\ctext{8}, {{\color[rgb]{1,0.4,0.1} orange}}&\cite{CMS:2015hmx}\\  \hline 
  $\tau\nu$ (Run\,2)&$y_Q \times y_\tau$& $m_{H^-}\ge400\,$GeV&\ctext{9}, {{\color[rgb]{0.4,0.4,0.4} grey}}&\cite{Sirunyan:2018lbg}\\  \hline \hline 
  ${{b\tau\nu}}$ (Run\,2)&$y_Q \times y_\tau$& all &\ctext{10}, {{\color[rgb]{0,0,0} black}}&---\\  \hline \hline 
   \end{tabular}
   }
    \caption{\label{Tab:cosnts} 
    Summary of the experimental constraints, relevant couplings and mass range, number in the figure and corresponding color.
    The current LHC bound is expressed by solid lines, and future prospects with 139$\ifb$, 500$\ifb$ and 3$\iab$ of  data are shown in dashed, dotted-dashed, dotted lines in the same color.
    We newly added the prospect of our $b\tau\nu$ signature with 139$\ifb$ of data.
}
\end{table}

Based on the background-only hypothesis, the upper limit on the event number $N^{95\%}$ is calculated at $95\%$ C.L. using the sum of the expected number of events in at least three $m_{\rm T}$ bins in a row, $N^{BG}$.
This procedure suppresses the effect of the statistical fluctuations in the distribution of the $Zjj$ BG category.  
We then subtract the BG event number in those bins, multiplied by a factor of 0.7, and obtain the maximum number of NP events,  $N_{\rm{NP max}}(=N^{95\%}-N^{BG}\times0.7)$.
Finally we deduce the NP sensitivity by comparing $N_{\rm NP max}$ and $0.7\times N_{\rm{NP}}$, where $N_{\rm{NP}}$ means the number of signal $H^-$ events in our simulation.

The resulting sensitivity assuming 139\,fb$^{-1}$ of  data is shown in Fig.\,\ref{Fig:coupling} by the black dashed line.
The dotted line denotes the HL-LHC sensitivity, assuming that the significance $\mathcal{S}$ scales as $\mathcal{S}\propto \sqrt{L}$, here $L$ denotes the integrated luminosity.
For the HL-LHC projection we assumed 3 ab$^{-1}$ of data.
We also show the various complementary experimental constraints following the color scheme in a previous paper \cite{Iguro:2022uzz}.

We find that the sensitivity of the $b\tau\nu$ signal almost covers the entire parameter region favored by the $R_{D^{(*)}}$ anomaly.
We also observe that it is easier to cover the heavier charged Higgs scenario: while the sensitivity of the $b\tau\nu$ search in the $(y_Q,y_\tau)$ plane depends only mildly on $m_{H^-}$, larger masses require larger couplings to solve the $R_{D^{(*)}}$ anomaly.
According to Eq.\,(\ref{Eq:Sig_Xs}), we see that the signal cross section is maximized at $|y_Q|=\sqrt{3} |y_\tau|$ when the product of couplings is fixed.
On the other hand, the cross section is minimized in the limit $|y_Q|\gg|y_\tau|$ thanks to the color factor in the normalization of the $H^-\to\tau\nu$ branching ratio.
As a result, the sensitivity is best around $|y_Q|\sim \sqrt{3} |y_\tau|$ and gets  worse for $|y_Q|\gg|y_\tau|$.

In the $m_{H^-}=180\,$GeV case, the combination with the existing low-mass di-jet search with $36\,\ifb$ of data and the $b\tau\nu$ prospect with $139\,\ifb$, corresponding to the moon symbol, is less constraining than the conservative bound from the $B_c\to\tau\nu$ decay.
However, once combined with the di-jet prospect for $139\,\ifb$, corresponding to the sun symbol, we can test a broader parameter space.
The HL-LHC reach denoted by the star symbol shows the great sensitivity and promising future of the $b\tau\nu$ channel. 

For all cases with $m_{H^-}\ge200\,$GeV we find an increased sensitivity, which grows with larger charged Higgs mass.
It is worth mentioning that the sensitivity of the $b\tau\nu$ signature is better than the $\tau\nu$ reach  even for $m_{H^-}\ge400\,$GeV.

For later convenience we define benchmark points in each figure which maximize the possible enhancement in $R_{D^{(*)}}$.
The numerical values of the Yukawa couplings $(y_Q,\,y_\tau)$ are listed in Table \ref{Tab:benchmarks}.
\begin{table}
\centering{
  \scalebox{0.97}{
\begin{tabular}{c||c|c|c}
\hline\hline
$m_{H^-}$ [GeV] & moon & sun & star \\\hline\hline
180 & (0.64,\,0.062) & (0.46,\,0.061) & (0.21,\,0.026) \\\hline
200 & (0.74,\,0.055) & (0.60,\,0.055) & (0.28,\,0.023) \\\hline
250 & (0.81,\,0.056) & (0.36,\,0.027) & -- \\\hline
300 & (0.88,\,0.061) & (0.47,\,0.026) & -- \\\hline
350 & (0.52,\,0.063) & -- & -- \\\hline
400 & (0.88,\,0.069) & -- & -- \\\hline\hline
\end{tabular}
}
}
\caption{Numerical values of the Yukawa couplings $(y_Q,y_\tau)$ for the benchmark points shown in Figs. \ref{Fig:coupling} and \ref{Fig:RD_sensitivity}.
\label{Tab:benchmarks}}
\end{table}

\subsection{Impact on the $H^-$ solution to the $R_{D^{(*)}}$ anomaly}

\begin{figure}[t]
\begin{center}
\includegraphics[scale=0.333]{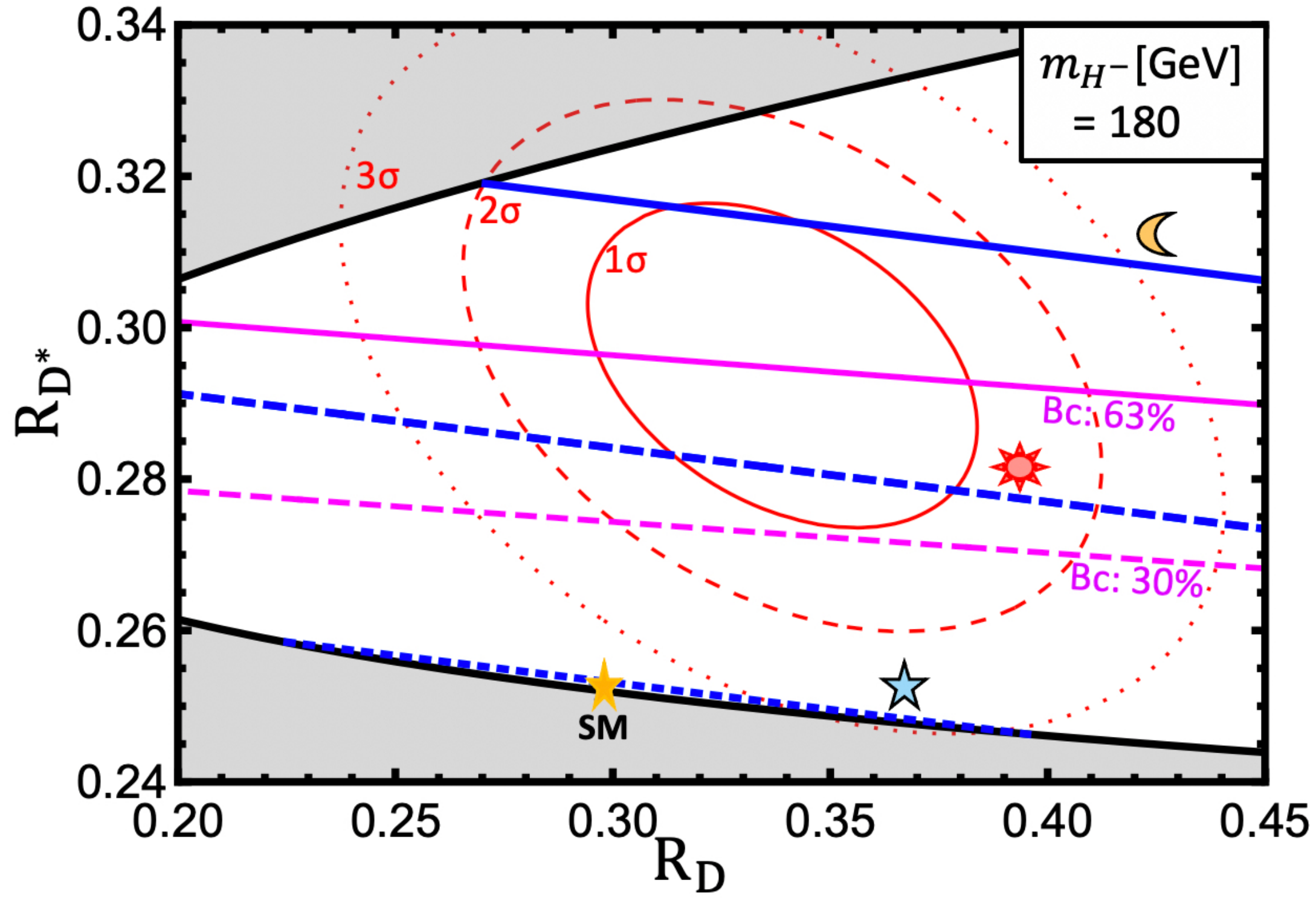}
\includegraphics[scale=0.333]{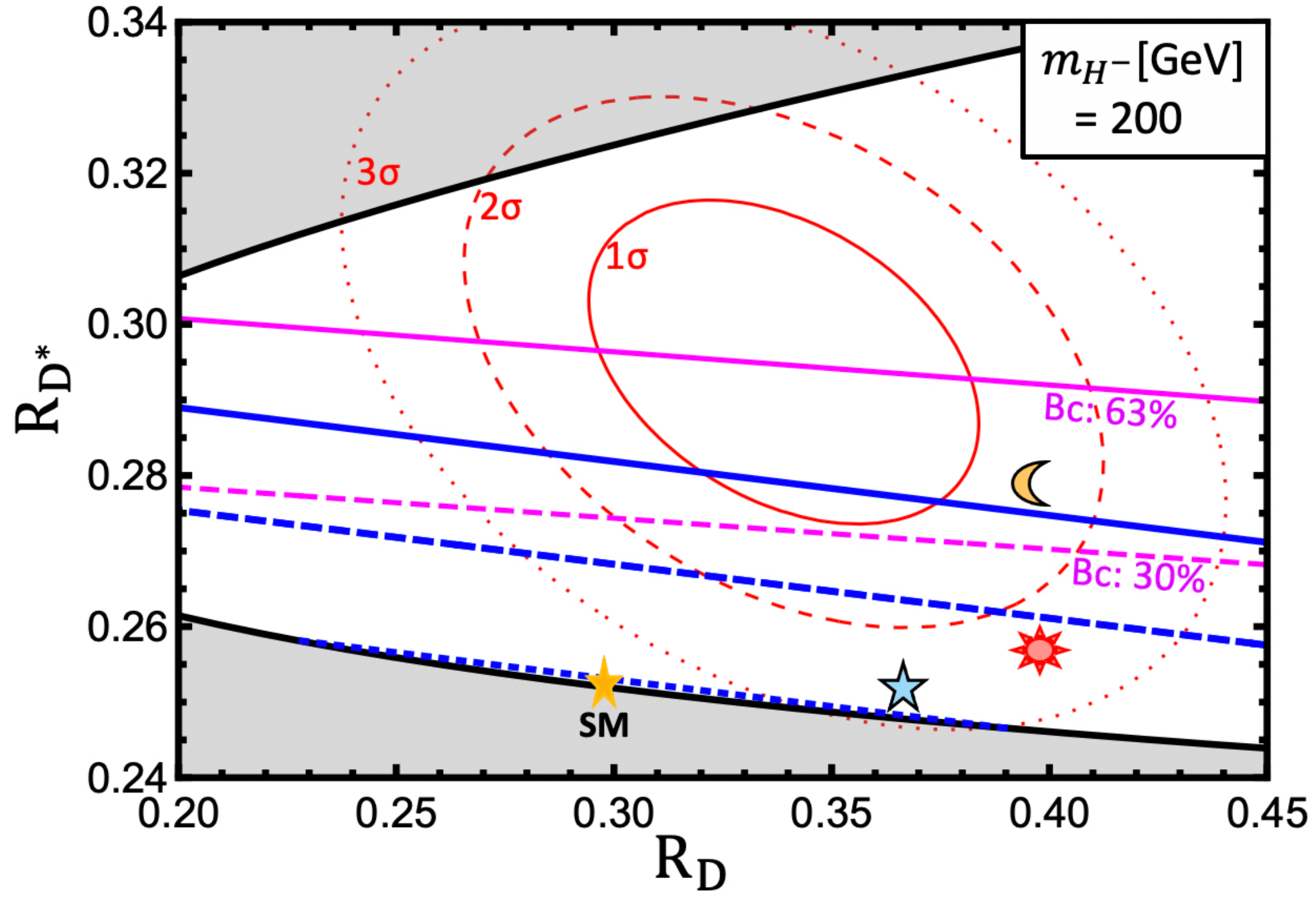}
\includegraphics[scale=0.333]{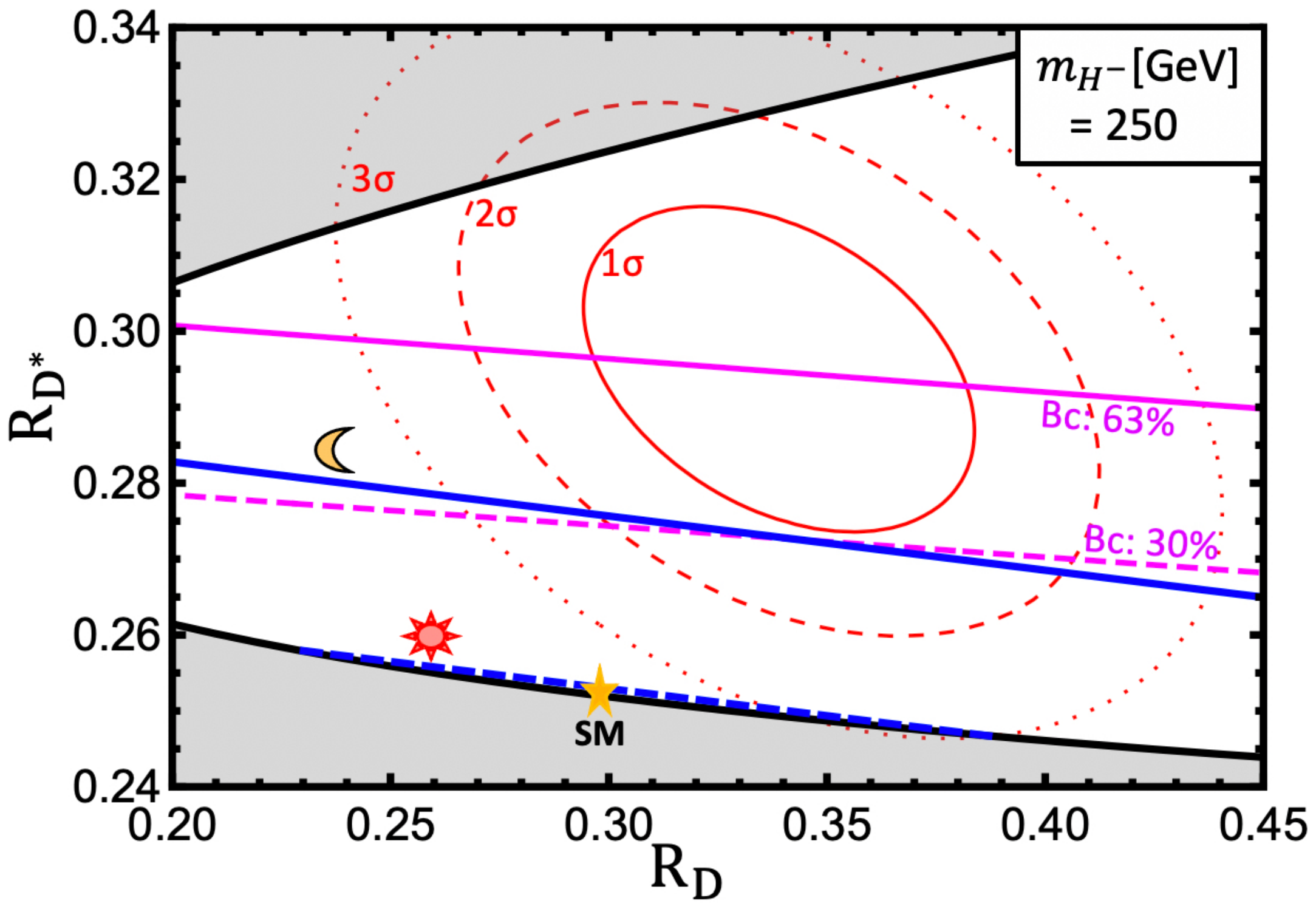}
\includegraphics[scale=0.333]{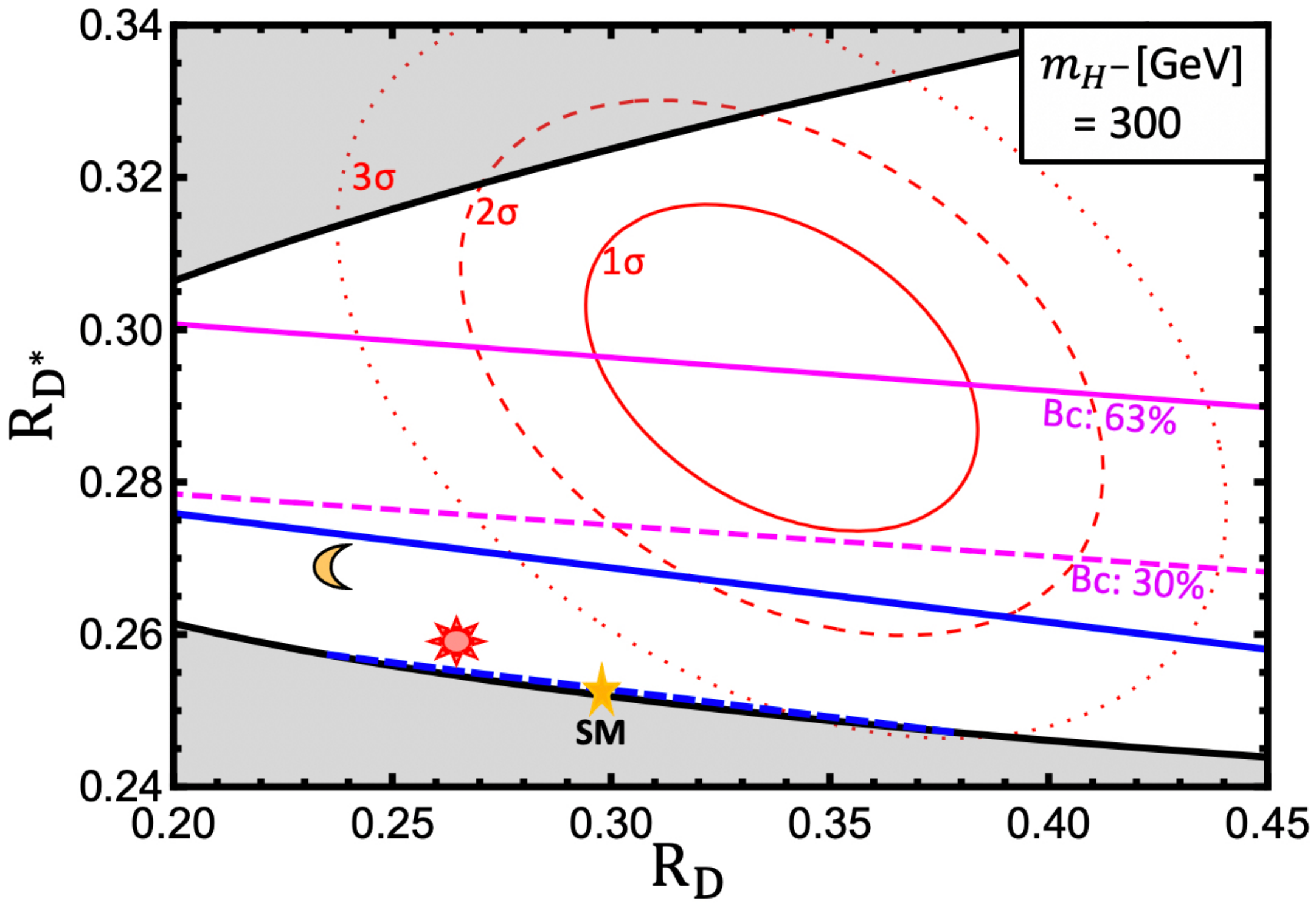}
\includegraphics[scale=0.333]{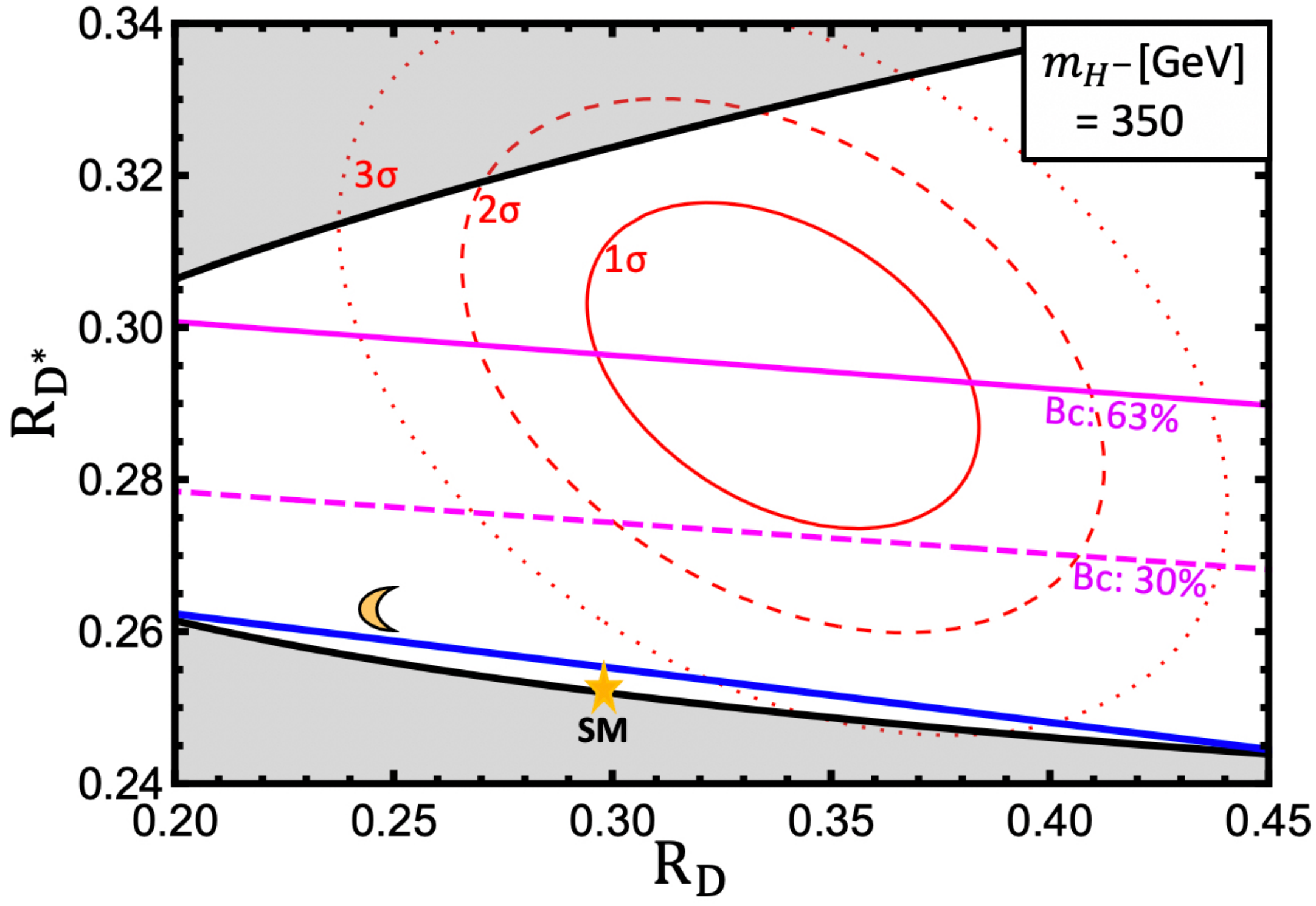}
\includegraphics[scale=0.333]{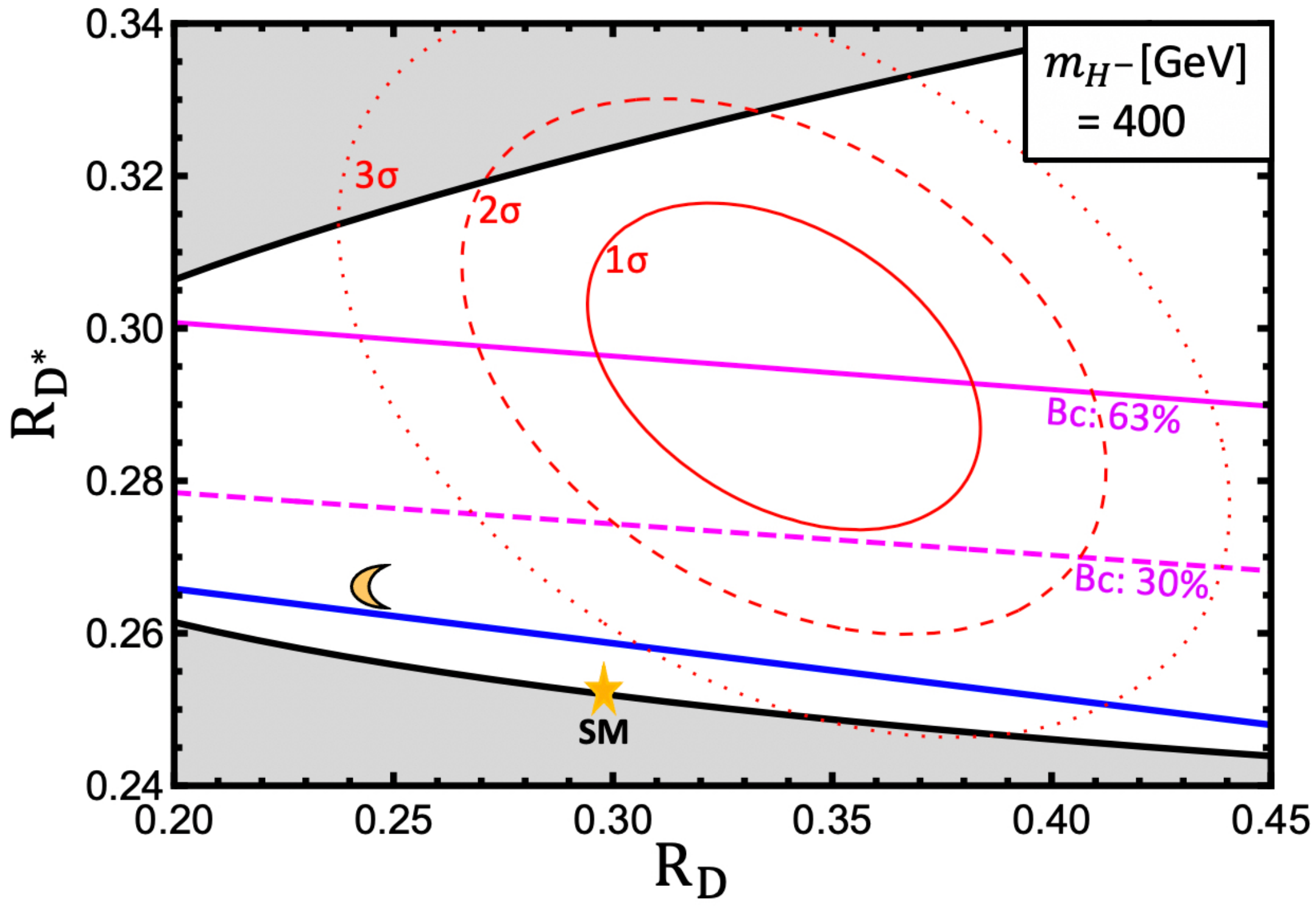}
\caption{
\label{Fig:RD_sensitivity}
Sensitivity of the $b\tau\nu$ channel shown in the $R_{D^{(*)}}$ plane, for fixed charged Higgs masses as shown in the upper right corner.
The blue lines show the predictions of the coupling combinations corresponding to the benchmark points in Fig.\,\ref{Fig:coupling} and Tab.\,\ref{Tab:benchmarks}, and are labeled by the respective moon, sun and star symbols. The area above the line is excluded by the respective bound.
The world average of the $R_{D^{(*)}}$ data at 1,\, 2 and 3\,$\sigma$ are shown by the red solid, dashed and dotted contours.
The HFLAV2021 SM prediction is indicated by a yellow star, and the horizontal magenta solid (dashed) line corresponds to BR($B_c\to\tau\nu)=63\,(30)\%$. 
The grey shaded region is not accessible within our model. 
} 
\end{center}
\end{figure}

In Fig.\,\ref{Fig:RD_sensitivity}, we project the sensitivity of the $b\tau\nu$ search to the $R_{D^{(*)}}$ plane.
To this end we show the $R_{D^{(*)}}$ predictions of the benchmark points defined in Fig.\,\ref{Fig:coupling} and Tab.\ \ref{Tab:benchmarks} that were chosen to maximize the enhancement in $R_{D^{(*)}}$.
Note that, in contrast to the LHC searches discussed above, the predictions for  $R_{D^{(*)}}$ are sensitive to the complex phases of the Yukawa couplings.
Therefore, by varying the complex phase, the benchmark points result in the predictions shown by the blue lines in the $R_{D^{(*)}}$ plane.
The red solid, dashed and dotted contours show the world average of the $R_{D^{(*)}}$ data at 1,\, 2 and 3\,$\sigma$.
The SM prediction shown as a yellow star is taken from HFLAV2021 \cite{Amhis:2019ckw}, and the horizontal magenta solid and dashed lines correspond to BR($B_c\to\tau\nu)=63$ and $30\,\%$. The area above the lines is exluded by the respective bound.
Note that the grey shaded region cannot be accessed within our model.

From Fig. \ref{Fig:RD_sensitivity} it is obvious that the $b\tau\nu$ signature provides a very powerful tool to test the low-mass charged Higgs interpretation of the $R_{D^{(*)}}$ anomaly.
For the entire charged-Higgs mass range, 139\,$\ifb$ of data provide an excellent sensitivity and can cover most of the $1\,\sigma$ range of the anomaly even for the most challenging case of $m_{H^-}=180\,$GeV.
For heavier charged-Higgs bosons, e.g. $m_{H^-}=300\,$GeV, the currently available data can even cover most of the $2\,\sigma$ region. 

In passing we note that selecting events with negatively charged $\tau$ leptons could further improve the sensitivity, as discussed in section\,\ref{Sec:strategy}. 
Furthermore, to suppress the dominant single top-originated BG, rejecting events with a large-$p_T$ $b$-jet could be a good option.
Finally we caution the reader that our evaluation is based on fast detector simulation, and further dedicated studies by the experimental collaborations are necessary to draw definite conclusions.

\section{\boldmath Conclusions}
\label{Sec:conclusion}
The current experimental data for the lepton-flavor universality ratios $R_{D^{(*)}}$ may imply the existence of new physics in $b\to c\tau\nu$ transitions.
Recently it was shown that a charged Higgs from a generic two Higgs doublet model can still explain the anomaly within $1\sigma$ when its mass is lighter than $400\,\GeV$.
Because of this low mass, it is expected that direct LHC searches can play an important role in testing this possibility, and the HL-LHC prospects have been assessed  in a previous paper~\cite{Iguro:2022uzz}.
There it was observed that it is difficult to test the whole range of the interesting parameter region based on extrapolations of the existing experimental results.

A $\tau\nu$ resonance search has been known to be a powerful tool to test the new physics effect in $b\to c\tau\nu$, however, it suffers from large SM background in the low $m_{\rm{T}}$ region.
An additional $b$-tagging can suppress this BG and improve the sensitivity, however it has not yet been performed by the experimental collaborations.
In this paper we studied the sensitivity of the $pp\to b H^\pm \to b\tau\nu$ signature to the low-mass region of the charged Higgs boson.

Our results show that most of the parameter region solving the $R_{D^{(*)}}$ anomaly can already be tested with the currently available LHC data.
If in a dedicated experimental $b\tau\nu$ search no excess is found, a major step towards ruling out the charged-scalar interpretation of the $R_{D^{(*)}}$ anomaly will be taken, favoring other new physics scenarios such as leptoquarks.

\section*{Acknowledgements}
\label{Sec:acknowledgement}
We would like to thank Joaquim Matias, Hiroyasu Yonaha and Ulrich Nierste for encouraging this project.
We also appreciate the support on computational resources from Martin Lang and Fabian Lange.
S.\,I. enjoys the support from the Japan Society for the Promotion of Science (JSPS) Core-to-Core Program, No.JPJSCCA20200002.
This work is also supported by the Deutsche Forschungsgemeinschaft (DFG, German Research Foundation) under grant 396021762-TRR\,257.


\bibliographystyle{utphys28mod}
\bibliography{ref}

\end{document}